\documentclass[journal]{IEEEtran}
\usepackage[table,xcdraw]{xcolor}
\usepackage{booktabs}
\usepackage{graphicx}
\usepackage{xcolor}
\usepackage{multirow}
\usepackage{threeparttable}
\usepackage{colortbl}
\usepackage[T1]{fontenc}
\usepackage[utf8]{inputenc}
\usepackage[bottom]{footmisc}
\usepackage{tablefootnote}
\usepackage{threeparttable}
\ifCLASSINFOpdf
\else
\fi
\begin{document}
\title{Computing Systems for Autonomous Driving: State-of-the-Art and Challenges
}
% \author{\textcolor{red}{\textbf{Deadline: August 15th}}}
% \author{Liangkai~Liu,~\IEEEmembership{Student Member,~IEEE,}
% Sidi Lu, Ren Zhong, Baofu Wu, Yongtao Yao, Qingyang Zhang,
%         Weisong~Shi,~\IEEEmembership{Fellow,~IEEE}}

\author{
\IEEEauthorblockN{
% {\textcolor{red}{REVISION DEADLINE: Nov. 15th 2020}} \\
% {\tt Technical Report: CAR-TR-2020-009}\\
Liangkai~Liu\IEEEauthorrefmark{1}\footnote{{\tt Technical Report: CAR-TR-2020-009}},
Sidi~Lu\IEEEauthorrefmark{1},
Ren~Zhong\IEEEauthorrefmark{1},
Baofu~Wu\IEEEauthorrefmark{1},
Yongtao~Yao\IEEEauthorrefmark{1},
Qingyang~Zhang\IEEEauthorrefmark{2},
Weisong~Shi\IEEEauthorrefmark{1}
}
\IEEEauthorblockA{
\\\IEEEauthorrefmark{1}Department of Computer Science, Wayne State University, Detroit, MI, USA, 48202\\
\IEEEauthorrefmark{2}School of Computer Science and Technology, Anhui University, Hefei, China, 230601\\
\{liangkai, lu.sidi, ren.zhong, baofu.wu, yongtaoyao, weisong\}@wayne.edu, qingyang@thecarlab.org
}
}

\maketitle

\begin{abstract}
\label{abstract}
% \liangkai{ The recent proliferation of computing technologies, e.g., sensors, computer vision, machine learning, hardware acceleration, and the wide deployment of communication mechanisms, e.g., DSRC, C-V2X, 5G, have pushed the horizon of autonomous driving, which automates the decision and control of vehicles by leveraging the perception results based on multiple sensors. The key to the success of these AV systems is making the reliable decision in a real-time fashion. 
% 3. What's the problem of AD? 4.high level goal of this paper. 5. what is included in this paper. }
The recent proliferation of computing technologies (e.g., sensors, computer vision, machine learning, and hardware acceleration), and the broad deployment of communication mechanisms (e.g., DSRC, C-V2X, 5G) have pushed the horizon of autonomous driving, which automates the decision and control of vehicles by leveraging the perception results based on multiple sensors. The key to the success of these autonomous systems is making a reliable decision in real-time fashion. However, accidents and fatalities caused by early deployed autonomous vehicles arise from time to time. The real traffic environment is too complicated for current autonomous driving computing systems to understand and handle. In this paper, we present state-of-the-art computing systems for autonomous driving, including seven performance metrics and nine key technologies, followed by twelve challenges to realize autonomous driving. We hope this paper will gain attention from both the computing and automotive communities and inspire more research in this direction.

% Due to the high safety and efficiency, autonomous driving has become the fundamental technology for the next generation of transportation. Although the autonomous driving vehicle has been built with rich sensors like camera, LiDAR, Radar, etc. and powerful computation accelerators like GPU, FPGA, DSP etc., it still fails in perception, sensing, and prediction frequently. To achieve level-4 or level-5 autonomous driving, the computing system of the autonomous driving vehicles still need huge improvement. In this paper, we tackle this from the computing system view. First we summarize the state-of-the-art works in computing systems for autonomous driving. Then we summarize eleven remaining challenges in the computing systems for full autonomous driving.
\end{abstract}

%\begin{IEEEkeywords}
%Autonomous driving, Computing %systems, Challenges.
%\end{IEEEkeywords}

% \fi
\IEEEpeerreviewmaketitle

\section{Introduction}
\label{introduction}

% Owing to the safety and efficiency in fuel consumption, autonomous driving techniques have attracted huge attention from both the academic and industry.

Recently, with the vast improvements in computing technologies, e.g., sensors, computer vision, machine learning, hardware acceleration, and the wide deployment of communication mechanisms, e.g., Dedicated short-range communications (DSRC), Cellular Vehicle-to-Everything (C-V2X), and 5G, autonomous driving techniques have attracted massive attention from both the academic and automotive communities~\cite{kenney2011dedicated,5g2016case}. According to~\cite{AVmarket}, the global autonomous driving market expects to grow up to \$173.15B by 2030. Many automotive companies have made enormous investments in this domain, including ArgoAI, Audi, Baidu, Cruise,  Mercedes-Benz, Tesla, Uber, and Waymo, to name a few~\cite{birdsall2014google,Mercedes-Benz,Audi-CES,somerville2018uber}. Several fleets of SAE (Society of Automotive Engineers) L4 vehicles have been deployed around the world~\cite{top-AV-companies,sae2018taxonomy}.

To achieve autonomous driving, determining how to make the vehicle understand the environment correctly and make safe controls in real-time is the essential task. Rich sensors including camera, LiDAR (Light Detection and Ranging), Radar, Inertial Measurement Unit (IMU), Global Navigation Satellite System (GNSS), and Sonar, as well as powerful computation devices, are installed on the vehicle~\cite{continental-radar,velodyne-lidar,beidou,gps,liu2017creating}. This design makes autonomous driving a real powerful "computer on wheels." In addition to hardware, the rapid development of deep learning algorithms in object/lane detection, simultaneous localization and mapping (SLAM), and vehicle control also promotes the real deployment and prototyping of autonomous vehicles~\cite{redmon2018yolov3,neven2018towards,thrun2008simultaneous,ziegler2014making}. The autonomous vehicle's computing systems are defined to cover everything (excluding the vehicle's mechanical parts), including sensors, computation, communication, storage, power management, and full-stack software. Plenty of algorithms and systems are designed to process sensor data and make a reliable decision in real-time. 

However, news of fatalities caused by early developed autonomous vehicles (AVs) arises from time to time. Until August 2020, five self-driving car fatalities happened for level-2 autonomous driving: four of them from Tesla and one from Uber~\cite{self-driving-fatality}. Table~\ref{tab:fatalities-list} summarizes the date, place, company, and reasons for these five fatalities. The first two fatalities attributed to Tesla happened in 2016 with the first accident occurring because neither the Autopilot system nor the driver failed to recognize the truck under thick haze. The vehicle in the second incident mistook the truck for open space. In the 2018 incident involving Tesla, the autopilot failed to recognize the highway divider and crashed into it. The most recent fatality from Tesla happened in 2019 because the vehicle failed to recognize a semi-trailer. The fatality from Uber happened because the autonomous driving system failed to recognize that pedestrians jaywalk. 

In summary, all four incidents associated with Tesla are due to perception failure, while Uber's incident happened because of the failure to predict human behavior. Another fact to pay attention to is that currently, the field-testing of level 2 autonomous driving vehicles mostly happens in places with good weather and light traffic conditions like Arizona and Florida. The real traffic environment is too complicated for the current autonomous driving systems to understand and handle easily. The objectives of level 4 and level 5 autonomous driving require colossal improvement of the computing systems for autonomous vehicles.

This paper presents state-of-the-art computing systems for autonomous driving, including seven performance metrics and nine key technologies, followed by eleven challenges and opportunities to realize autonomous driving. The remaining parts of this paper are organized as follows: Section~\ref{reference-arch} discusses the reference architecture of the computing systems for autonomous driving. In Section~\ref{metrics}, we
show some metrics used in the evaluation of the computing system. Section~\ref{key-tech} discusses the key technologies for autonomous driving. Section~\ref{challenges} presents possible challenges. Finally, this paper concludes in Section~\ref{conclusion}.

% In summary, this paper makes the following contributions: 
% % \liangkai{need more details and conclusions for each contribution. Not just what you have done. }
% \begin{itemize}
%     \item We 
%     The time variation issues in DNN inference for autonomous driving are thoroughly studied.
%     \item Through a , we observed .
% \end{itemize}

% \usepackage{booktabs}
\begin{table*}[ht]
\centering
\caption{List of fatalities caused by Level 2 autonomous driving vehicles.}
\label{tab:fatalities-list}
\begin{tabular}{@{}cccc@{}}
\toprule
\textbf{Date} & \textbf{Place} & \textbf{Company} & \textbf{Reason} \\ \midrule
20 Jan. 2016 & Handan, Hebei China & Tesla & fail to recognize truck under a thick haze \\
07 May 2016 & Williston, Florida USA & Tesla & mistook the truck for open sky \\
18 Mar. 2018 & Tempe, Arizona USA & Uber & fail to recognize pedestrians jaywalk at night \\
23 Mar. 2018 & Mountain View, California USA & Tesla & fail to recognize the highway divider \\
1 Mar. 2019 & Delray Beach, Florida USA & Tesla & fail to recognize semi-trailer \\ \bottomrule
\end{tabular}
\end{table*}

\section{Reference Architecture}
\label{reference-arch}
% Reference architecture, figure to show the computing system

As an essential part of the whole autonomous driving vehicle, the computing system plays a significant role in the whole pipeline of driving autonomously. There are two types of designs for computing systems on autonomous vehicles: modular-based and end-to-end based. 

Modular design decouples the localization, perception, control, etc. as separate modules and make it possible for people with different backgrounds to work together~\cite{kato2015open}. The DARPA challenges is a milestone for the prototyping of autonomous driving vehicles, including Boss from CMU~\cite{urmson2008autonomous}, Junior from Stanford~\cite{levinson2011towards}, TerraMax and BRAiVE from University of Parma~\cite{broggi2013extensive}, etc. Their designs are all based on modules including perception, mission planning, motion planning, and vehicle controls. Similarly, the survey fleet vehicles developed by Google and Uber are also modular-based~\cite{birdsall2014google,somerville2018uber}. The main differences for these AV prototypes are the software and the configuration of sensors like camera, LiDAR, Radar, etc.

In contrast, the end-to-end based design is largely motivated by the development of artificial intelligence. Compared with modular design, end-to-end system purely relies on machine learning techniques to process the sensor data and generate control commands to the vehicle~\cite{muller2006off,chen2015deepdriving,bojarski2016end,xu2017end,sallab2017deep,kendall2019learning}. Table~\ref{tab:architecture-design} shows a detailed description of these end-to-end designs. Four of them are based on supervised DNNs to learn driving patterns and behaviors from human drivers. The remaining two are based on Deep Q-Network (DQN), which learns to find the optimum driving by itself. Although the end-to-end based approach promises to decrease the modular design's error propagation and computation complexity, there is no real deployment and testing of it~\cite{ADsurvey2020}.

\begin{table}[]
\caption{End-to-end approaches for autonomous driving.}
\label{tab:architecture-design}
\begin{tabular}{@{}ccl@{}}
\toprule
\textbf{Work} & \textbf{Methods} & \multicolumn{1}{c}{\textbf{Characteristics}} \\ \midrule
\cite{muller2006off} & supervised DNN & \begin{tabular}[c]{@{}l@{}}raw image to steering angles for off-road \\ obstacle avoidance on mobile robots\end{tabular} \\
\rowcolor[HTML]{EFEFEF} 
\cite{chen2015deepdriving} & supervised DNN & \begin{tabular}[c]{@{}l@{}}map an input image to a small number of \\ key perception indicators\end{tabular} \\
\cite{bojarski2016end} & supervised DNN & \begin{tabular}[c]{@{}l@{}}CNN to map raw pixels from a camera \\ directly to steering commands\end{tabular} \\
\rowcolor[HTML]{EFEFEF} 
\cite{xu2017end} & supervised DNN & \begin{tabular}[c]{@{}l@{}}FCN-LSTM network to predict multi-modal \\ discrete and continuous driving behaviors\end{tabular} \\
\cite{sallab2017deep} & DQN & \begin{tabular}[c]{@{}l@{}}automated driving framework in simulator \\ environment\end{tabular} \\
\rowcolor[HTML]{EFEFEF} 
\cite{kendall2019learning} & DQN & \begin{tabular}[c]{@{}l@{}}lane following in a countryside road without \\ traffic using a monocular image as input \end{tabular} \\ \bottomrule
\end{tabular}
\end{table}

\begin{figure}[!htp]
	\centering
	\includegraphics[width=\columnwidth]{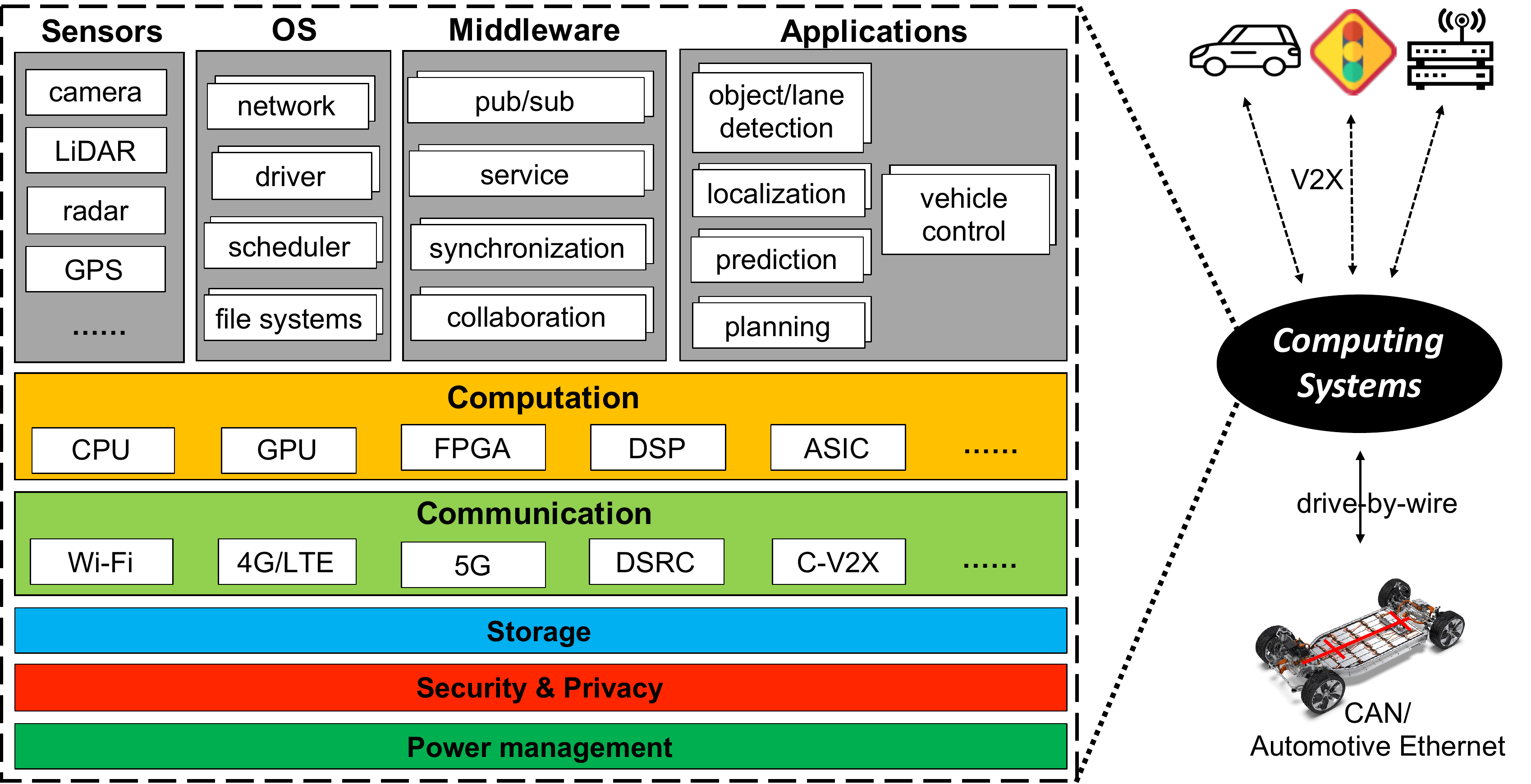}
	\caption{Representative reference architecture of the computing system for autonomous driving.}
	\label{fig:framework}
\end{figure}

As most prototypes are still modular-based, we choose it as the basis for the computing system reference architecture. Figure~\ref{fig:framework} shows a representative reference architecture of the computing system on autonomous vehicles. Generally, the computing system for autonomous driving vehicles can be divided into computation, communication, storage, security and privacy, and power management. Each part covers four layers with sensors, operating system (OS), middleware, and applications. The following paragraphs will discuss the corresponding components.

% The main components of the computing system includes hardware devices: sensors, computation device, communication device, storage, and power management; software systems: operating system (OS), middleware system, applications, and security and privacy system. 

% \liangkai{please explain the main components of the figure here. The following paragraphs will discuss corresponding components.}

For safety, one of the essential tasks is to enable the ``computer'' to understand the road environment and send correct control messages to the vehicle. The whole pipeline starts with the sensors. Plenty of sensors can be found on an autonomous driving vehicle: camera, LiDAR, radar, GPS/GNSS, ultrasonic, inertial measurement unit (IMU), etc. These sensors capture real-time environment information for the computing system, like the eyes of human beings. Operating system (OS) plays a vital role between hardware devices (sensors, computation, communication) and applications. Within the OS, drivers are bridges between the software and hardware devices; the network module provides the abstraction communication interface; the scheduler manages the competition to all the resources; the file system provides the abstraction to all the resources. For safety-critical scenarios, the operating system must satisfy real-time requirements.

As the middle layer between applications and operating systems~\cite{schantz2002middleware}, middleware provides usability and programmability to develop and improve systems more effectively. Generally, middleware supports publish/subscriber, remote procedure call (RPC) or service, time synchronization, and multi-sensor collaboration. A typical example of the middleware system is the Robot Operating System (ROS)~\cite{quigley2009ros}. 
On top of the operating system and middleware system, several applications, including object/lane detection, SLAM, prediction, planning, and vehicle control, are implemented to generate control commands and send them to the vehicle's drive-by-wire system. Inside the vehicle, several Electronic Control Units (ECUs)  are used to control the brake, steering, etc., which are connected via Controller Area Network (CAN bus) or Automotive Ethernet~\cite{hank2013automotive}. In addition to processing the data from on-board sensors, the autonomous driving vehicle is also supposed to communicate with other vehicles, traffic infrastructures, pedestrians, etc. as complementary.

\section{Metrics for Computing System}
\label{metrics}

% metrics for computing system: accuracy, power, cost (system), timeliness, privacy and security

According to the report about autonomous driving technology from the National Science \& Technology Council (NSTC) and the United States Department of Transportation (USDOT)~\cite{ensuring-USDoT}, ten technology principles are designed to foster research, development, and integration of AVs and guide consistent policy across the U.S. Government. These principles cover safety, security, cyber security, privacy, data security, mobility, accessibility, etc. Corresponding to the autonomous driving principles, we define several metrics to evaluate the computing system's effectiveness.

% As a safety critical system, the efficiency and reliability of the autonomous vehicles' computing system are of the utmost importance 

\textbf{Accuracy}
Accuracy is defined to evaluate the difference between the detected/processed results with the ground truth. Take object detection and lane detection, for example, the Intersection Over Union (IOU) and mean Average Precision (mAP) are used to calculate the exact difference between the detected bounding box of objects/lanes and the real positions~\cite{everingham2010pascal,girshick2014rich}. For vehicle controls, the accuracy would be the difference between the expected controls in break/steering with the vehicle's real controls. 

\textbf{Timeliness}
Safety is always the highest priority. Autonomous driving vehicles should be able to control themselves autonomously in real-time. According to~\cite{kato2015open}, if the vehicle is self-driving at 40km per hour in an urban area and wants the control effective every 1 meter, then the whole pipeline's desired response time should be less than 90ms. To satisfy the desired response time, we need each module in the computing system to finish before the deadline.  

\textbf{Power}
Since the on-board battery powers the whole computing system, the computing system's power dissipation can be a big issue. For electrical vehicles, the computing system's power dissipation for autonomous driving reduces the vehicle's mileage with up to 30\%~\cite{energy-reduce}. In addition to mileage, heat dissipation is another issue caused by high power usage. Currently, the NVIDIA Drive PX Pegasus provides 320 INT8 TOPS of AI computational power with a 500 watts budget~\cite{nvidia-xavier}. With the power budget of sensors, communication devices, etc., the total power dissipation will be higher than 1000 watts. The power budget is supposed to be a significant obstacle for producing the real autonomous driving vehicle.

\textbf{Cost}
Cost is one of the essential factors that affect the board deployment of autonomous vehicles. According to~\cite{AV-cost,Waymo-lidar}, the cost of a level 4 autonomous driving vehicle attains 300,000 dollars, in which the sensors, computing device, and communication device cost almost 200,000 dollars. In addition to the hardware cost, the operator training and vehicle maintenance cost of AVs (like insurance, parking, and repair) is also more expensive than traditional vehicles.

\textbf{Reliability}
To guarantee the safety of the vehicle, reliability is a big concern~\cite{sabaliauskaite2018integrating}. On one hand, the worst-case execution time is supposed to be longer than the deadline. Interruptions or emergency stops should be applied in such cases. On the other hand, failures happen in sensors, computing/communication devices, algorithms, and systems integration~\cite{orrick1994failure}. How to handle these potential failures is also an essential part of the design of the computing system. 

\textbf{Privacy}
As the vehicle captures a massive amount of sensor data from the environment, vehicle data privacy becomes a big issue. For example, the pedestrian's face and the license plate captured by the vehicle's camera should be masked as soon as possible~\cite{samira-edgemask}. Furthermore, who owns the driving data is also an important issue, which requires the system's support for data access, storage, and communication~\cite{attackPIEEE}. 

\textbf{Security}
% security level
The secureness of the on-board computing system is essential to the success of autonomous driving since, ultimately, the computing system is responsible for the driving process. Cyber attacks can be launched quickly to any part of the computing system~\cite{attackPIEEE, sensorattacks}. We divide the security into four aspects: sensing security, communication security, data security, and control security~\cite{yan2016sensorattack,gpsattack}. We envision that the on-board computing system will have to pass a certain security test level before deploying it into real products.  

\section{Key Technologies}
\label{key-tech}

% 3 pages
% \textcolor{red}{\textbf{each subsection half page (one column)}}
% \liangkai{please add a short connecting paragraph here.}

\begin{figure}[!htp]
	\centering
	\includegraphics[width=\columnwidth]{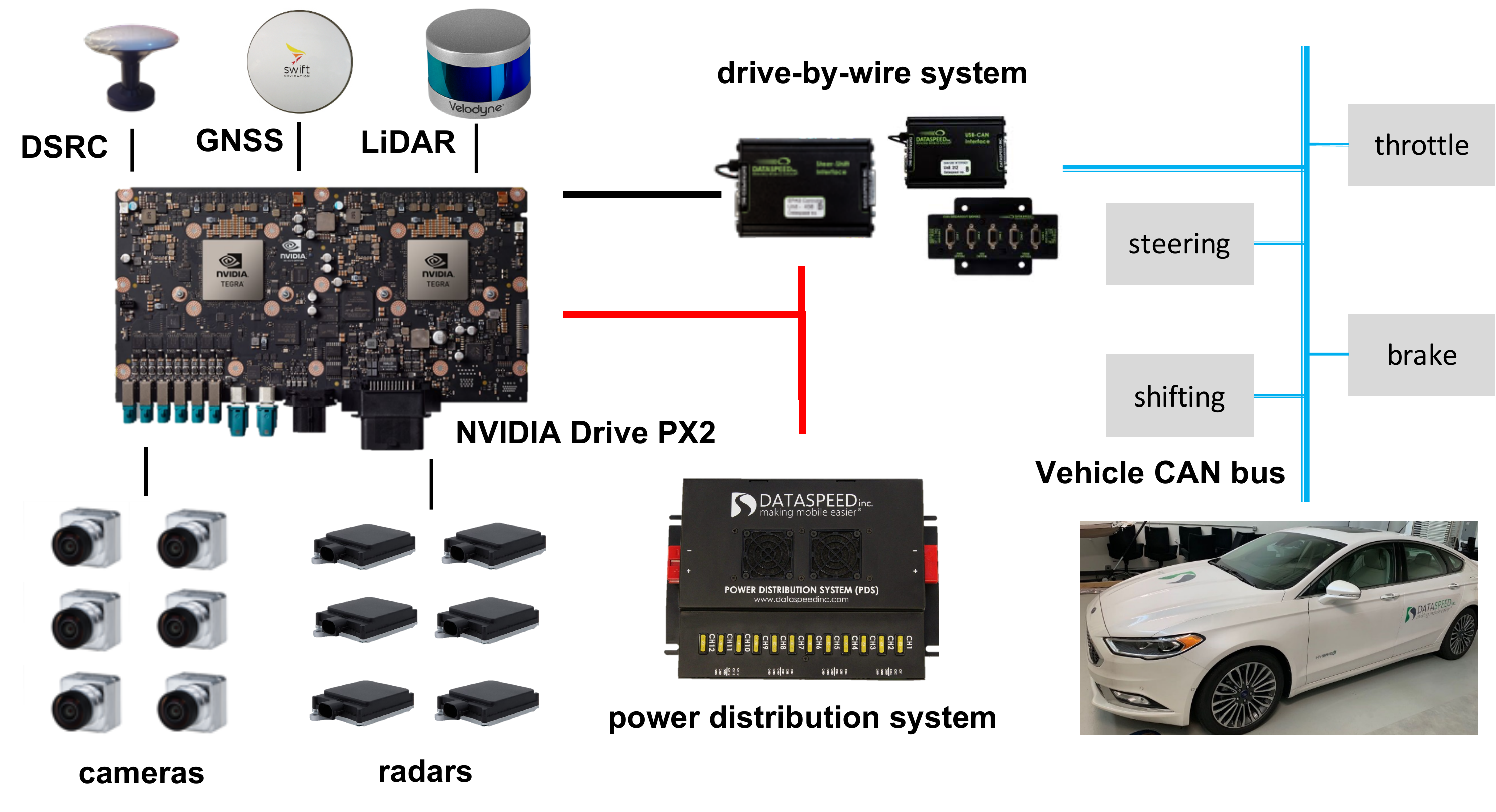}
	\caption{A typical example of a computing system for autonomous driving.}
	\label{fig:typical-example}
\end{figure}

An autonomous vehicle involves multiple subjects, including computing systems, machine learning, communication, robotics, mechanical engineering, and systems engineering, to integrate different technologies and innovations. Figure~\ref{fig:typical-example} shows a typical example of autonomous driving vehicles called \textit{Hydra}, which is developed by The CAR lab at Wayne State University~\cite{carlab}. An NVIDIA Drive PX2 is used as the vehicle computation unit (VCU). Multiple sensors, including six cameras, six radars, one LiDAR, one GNSS antenna, and one DSRC antenna, are installed for sensing and connected with VCU.  The CAN bus is used to transmit messages between different ECUs controlling steering, throttle, shifting, brake, etc. Between the NVIDIA Drive PX2 and the vehicle's CAN bus, a drive-by-wire system is deployed as an actuator of the vehicle control commands from the computing system. Additionally, a power distribution system is used to provide extra power for the computing system. It is worth noting that the computing system's power distribution is non-negligible in modern AVs~\cite{liu2019e2m}. In this section, we summarize several key technologies and discuss their state-of-the-art.

\subsection{Sensors}

\subsubsection{Cameras}
In terms of usability and cost, cameras are the most popular sensors on autonomous driving vehicles. The camera image gives straightforward 2D information, making it useful in some tasks like object classification and lane tracking. Also, the range of the camera can vary from several centimeters to near one hundred meters. The relatively low cost and commercialization production also contribute to the complete deployment in the real autonomous driving vehicle. However, based on lights, the camera's image can be affected by low lighting or bad weather conditions. The usability of the camera decreases significantly under heavy fog, raining, and snowing. Futhermore, the data from the camera is also a big problem. On average, every second, one camera can produce 20-40MB of data. %How to process and store these data can be a big challenge.

\subsubsection{Radar}
The radar's full name is Radio Detection and Ranging, which means to detect and get the distance using radio. The radar technique measures the Time of Flight (TOF) and calculates the distance and speed. Generally, the working frequency of the vehicle radar system is 24GHz or 77GHz. Compared with 24GHz, 77GHz shows higher accuracy in distance and speed detection. Besides, 77GHz has a smaller antenna size, and it has less interference than 24GHz. For 24GHz radar, the maximum detection range is 70 meters, while the maximum range increases to 200 meters for 77GHz radar. According to~\cite{continental-radar}, the price for Continental's long-range radar can be around \$3000, which is higher than the camera's price. However, compared with a camera, radar is less affected by the weather and low lighting environment, making it very useful in some applications like object detection and distance estimation. The data size is also smaller than the camera. Each radar produces 10-100KB per second.

\subsubsection{LiDAR}
Similar to Radar, LiDAR's distance information is also calculated based on the TOF. The difference is that LiDAR uses the laser for scanning, while radar uses electromagnetic waves. LiDAR consists of a laser generator and a high accuracy laser receiver. LiDAR generates a three-dimensional image of objects, so it is widely used to detect static objects and moving objects. LiDAR shows good performance with a range from several centimeters to 200 meters, and the accuracy of distance goes to centimeter-level. LiDAR is widely used in object detection, distance estimation, edge detection, SLAM~\cite{zhang2018pirvs,thrun2008simultaneous}, and High-Definition (HD) Map generation~\cite{levinson2007map,konolige2004large,fairfield2011traffic,ziegler2014making}. Compared with the camera, LiDAR shows larger sensing range and its performance is less affected by bad weather and low lighting. However, in terms of the cost, LiDAR seems less competitive than camera and radar. According to~\cite{velodyne-lidar}, the 16 lines Velodyne LiDAR costs almost \$8000, while the Velodyne VLS-128E costs over \$100,000. High costs restrict the wide deployment of LiDAR on autonomous vehicles, contributing to the autonomous vehicle's high cost. LiDAR can generate almost 10-70MB data per second, a huge amount of data for the computing platform to process in real-time. 

\subsubsection{Ultrasonic sensor}
Ultrasonic sensor is based on ultrasound to detect the distance. Ultrasound is a particular sound that has a frequency higher than 20kHz. The distance is also detected by measuring TOF. The ultrasonic sensor's data size is close to the radar's, which is 10-100KB per second. Besides, the ultrasonic sensor shows good performance in bad weather and low lighting environment. The ultrasonic sensor is much cheaper than the camera and radar. The price of the ultrasonic sensor is always less than \$100. The shortcoming of ultrasonic sensors is the maximum range of only 20 meters, limiting its application to short-range detection like parking assistance.

\begin{table*}
\centering
\caption{Comparisons of camera, radar, LiDAR, and ultrasonic sensor.}
\label{tab:sensor-comparison}
\begin{tabular}{cccccc} 
\hline
 \textbf{Metrics}  & \textbf{Human}  & \textbf{Camera}  & \textbf{Radar}  & \textbf{LiDAR}  & \textbf{Ultrasonic}  \\ 
\hline
Techniques & - & Lights & Electromagnetic & Laser Reflection & Ultrasound \\
\rowcolor[rgb]{0.839,0.839,0.839} Sensing Range & 0-200m & 0-100m & \begin{tabular}[c]{@{}>{\cellcolor[rgb]{0.839,0.839,0.839}}c@{}}1cm-200m (77GHz)\\ 1cm-70m (24GHz) \end{tabular} & 0.7-200m & 0-20m \\
Cost & - & $\sim$\$500 & $\sim$\$3,000 & \$5,000 - \$100,000 & $\sim$\$100 \\
\rowcolor[rgb]{0.839,0.839,0.839} Data per second & - & 20-40MB & 10-100KB & 10-70MB & 10-100KB \\
\begin{tabular}[c]{@{}c@{}} Bad weather \\ functionality \end{tabular} & Fair & Poor & Good & Fair & Good \\
\rowcolor[rgb]{0.839,0.839,0.839} \begin{tabular}[c]{@{}>{\cellcolor[rgb]{0.839,0.839,0.839}}c@{}}Low lighting \\ functionality \end{tabular} & Poor & Fair & Good & Good & Good \\
\begin{tabular}[c]{@{}c@{}} Application \\ Scenarios \end{tabular} & \begin{tabular}[c]{@{}c@{}}Object Detection\\ Object Classification\\ Edge Detection\\ Lane Tracking \end{tabular} & \begin{tabular}[c]{@{}c@{}}Object Classification\\ Edge Detection\\ Lane Tracking \end{tabular} & \begin{tabular}[c]{@{}c@{}}Object Detection\\ Distance Estimation \end{tabular} & \begin{tabular}[c]{@{}c@{}}Object Detection\\ Distance Estimation\\ Edge Detection \end{tabular} & \begin{tabular}[c]{@{}c@{}}Object Detection\\ Distance Estimation \end{tabular} \\
\hline
\end{tabular}
\end{table*}

\subsubsection{GPS/GNSS/IMU}
Except for sensing and perception of the surrounding environment, localization is also a significant task running on top of the autonomous driving system. In the localization system of the autonomous vehicle, GPS, GNSS, and IMU are widely deployed. GNSS is the name for all the satellite navigation systems, including GPS developed by the US, Galileo from Europe, and BeiDou Navigation Satellite System (BDS)~\cite{beidou} from China. The accuracy of GPS can vary from several centimeters to several meters when different observation values and different processing algorithms are applied~\cite{gps}. The strengths of GPS are low costs, and the non-accumulation of error over time. The drawback of GPS is that the GPS deployed on current vehicles only has accuracy within one meter: and GPS requires an unobstructed view in the sky, so it does not work in environments like tunnels, for example. Besides, the GPS sensing data updates every 100ms, which is not enough for the vehicle's real-time localization.

% \subsubsection{IMU}
IMU stands for inertial measurement unit, which consists of gyroscopes and accelerometers. Gyroscopes are used to measure the axes' angular speed to calculate the carrier's position. In comparison, the accelerometer measures the object's three axes' linear acceleration and can be used to calculate the carrier's speed and position. The strength of IMU is that it does not require an unobstructed view from the sky. The drawback is that the accuracy is low, and the error is accumulated with time. IMU can be a complementary sensor to the GPS because it has an updated value every 5ms, and it works appropriately in environments like tunnels. Usually, a Kalman filter is applied to combine the sensing data from GPS and IMU to get fast and accurate localization results~\cite{liu2017creating}.

Table~\ref{tab:sensor-comparison} shows a comparison of sensors, including camera, radar, LiDAR, and ultrasonic sensors with human beings. From the comparison, we can easily conclude that although humans have strength in the sensing range and show more advantaged application scenarios than any sensor, the combination of all the sensors can do a better job than human beings, especially in bad weather and low lighting conditions.

\subsection{Data Source}

\subsubsection{Data characteristics}
% Qingyang AC4AV
As we listed before, various sensors, such as GPS, IMU, camera, LiDAR, radar, are equipped in AVs, and they will generate hundreds of megabytes of data per second, fed to different autonomous driving algorithms. 
The data in AVs could be classified into two categories: real-time data, and historical data. Typically, the former is transmitted by a messaging system with the Pub/Sub pattern in most AVs solutions, enabling different applications to access one data simultaneously. 
Historical data includes application data. The data persisted from real-time data, where structured data, i.e., GPS, is stored into a database, and unstructured data, i.e., video, is stored as files. 
% In this case, the data management and permission management (or say access control) are more complex than traditional systems, which only need to face one category of real-time data or historical data. 

\subsubsection{Dataset and Benchmark}
Autonomous driving dataset is collected by survey fleet vehicles driving on the road, which provides the training data for research in machine learning, computer vision, and vehicle control. Several popular datasets provide benchmarks, which are rather useful in autonomous driving systems and algorithms design.
% , giving researchers the ability to measure and make tradeoffs in architectural and algorithmic decisions. 
% The main challenge of making a dataset is how to collect as much as possible the complicated surrounding environment and challenging scenarios, such as rain, snow and accident. 
Here are a few popular datasets: (1)~\textit{KITTI}: As one of the most famous autonomous driving dataset, the KITTI~\cite{geiger2013vision} dataset covers stereo, optical flow, visual odometry, 3D object detection, and 3D tracking. It provides several benchmarks, such as stereo, flow, scene, optical flow, depth, odometry, object tracking~\cite{hong2016online}, road, and semantics~\cite{garcia2017review}. (2)~\textit{Cityscapes}: For the semantic understanding of urban street scenes, the Cityscapes \cite{tampuu2020survey} dataset includes 2D semantic segmentation on pixel-level, instance-level, and panoptic semantic labeling, and provides corresponding benchmarks on them. (3)~\textit{BDD100K}: As a large-scale and diverse driving video database, BDD100K~\cite{yu2020bdd100k} consists of 100,000 videos and covers different weather conditions and times of the day. (4) ~\textit{DDD17}: As the first end-to-end dynamic and active-pixel vision sensors (DAVIS) driving dataset, DDD17~\cite{binas2017ddd17} has more than 12 hours of DAVIS sensor data under different scenarios and different weather conditions, as well as vehicle control information like steering, throttle, and brake.

\subsubsection{Labeling}
Data labeling is an essential step in a supervised machine learning task, and the quality of the training data determines the quality of the model. Here are a few different types of annotation methods: (1) \textit{Bounding boxes}: the most commonly used annotation method (rectangular boxes) in object detection tasks to define the location of the target object, which can be determined by the x and y-axis coordinates in the upper-left corner and the lower-right corner of the rectangle. (2) \textit{Polygonal segmentation}: since objects are not always rectangular, polygonal segmentation is another annotation approach where complex polygons are used to define the object's shape and location in a considerably precise way. (3) \textit{Semantic segmentation}: a pixel-wise annotation, where every pixel in an image is assigned to a class. It is primarily used in cases where environmental context is essential. (4) \textit{3D cuboids}: They provide 3D representations of the objects, allowing models to distinguish features like volume and position in a 3D space. (5) \textit{Key-Point and Landmark} are used to detect small objects and shape variations by creating dots across the image. As to the annotation software, MakeSense.AI~\cite{makesense}, LabelImg~\cite{labelImg}, 
VGG image annotator~\cite{annotator}, LabelMe~\cite{LabelMe}, Scalable~\cite{Scalable}, and RectLabel~\cite{RectLabel} are popular image annotation tools.

\subsection{Autonomous Driving Applications}
% \liangkai{need add one overview sentence here.}
Plenty of algorithms are deployed in the computing system for sensing, perception, localization, prediction, and control. In this part, we present the state-of-the-art works for algorithms including object detection, lane detection, localization and mapping, prediction and planning, and vehicle control.

\subsubsection{Object detection} Accurate object detection under challenging scenarios is essential for real-world deep learning applications for AVs~\cite{michaelis2019benchmarking}. In general, it is widely accepted that the development of object detection algorithms has gone through two typical phases: (1) conventional object detection phase, and (2) deep learning supported object detection phase~\cite{zou2019object}. Viola Jones Detectors~\cite{viola2001rapid}, Histogram of Oriented Gradients (HOG) feature descriptor~\cite{dalal2005histograms}, and Deformable Part-based Model (DPM)~\cite{felzenszwalb2008discriminatively} are all the typical traditional object detection algorithms. Although today’s most advanced approaches have far exceeded the accuracy of traditional methods, many dominant algorithms are still deeply affected by their valuable insights, such as hybrid models, bounding box regression, etc. As to the deep learning-based object detection approaches, the state-of-the-art methods include the Regions with CNN features (RCNN) series~\cite{girshick2015fast, ren2015faster, girshick2014rich, he2017mask}, Single Shot MultiBox Detector (SSD) series~\cite{liu2016ssd, fu2017dssd}, and You Only Look Once (YOLO) series~\cite{redmon2016you, redmon2017yolo9000, redmon2018yolov3}. Girshick et al. first introduced deep learning into the object detection field by proposing RCNN in 2014~\cite{girshick2014rich, girshick2015region}. Later on, Fast RCNN~\cite{girshick2015fast} and Faster RCNN~\cite{ren2015faster} were developed to accelerate detection speed. In 2015, the first one-stage object detector, i.e., YOLO was proposed~\cite{redmon2016you}. Since then,  the YOLO series algorithms have been continuously proposed and improved, for example, YOLOv3~\cite{redmon2018yolov3} is one of the most popular approaches, and YOLOv4~\cite{bochkovskiy2020yolov4} is the latest version of the YOLO series. To solve the trade-off problem between speed and accuracy, Liu et al. proposed SSD~\cite{liu2016ssd} in 2015, which introduces the regression technologies for object detection. Then, RetinaNet was proposed in 2017~\cite{lin2017focal} to further improve detection accuracy by introducing a new loss function to reshape the standard cross-entropy loss.

% From the application point of view, object detection can be grouped into two research topics “general object detection” and “detection applications”, where the former one aims to explore the methods of detecting different types of objects under a unified framework to simulate the human vision and cognition, and the later one refers to the detection under specific application scenarios, such as pedestrian detection, face detection, text detection, etc.

\subsubsection{Lane detection}
Performing accurate lane detection in real-time is a crucial function of advanced driver-assistance systems (ADAS)~\cite{neven2018towards}, since it enables AVs to drive themselves within the road lanes correctly to avoid collisions, and it supports the subsequent trajectory planning decision and lane departure. 

Traditional lane detection approaches (e.g. \cite{borkar2011novel, deusch2012random, hur2013multi, jung2013efficient, tan2014novel, wu2014lane}) aims to detect lane segments based on diverse handcrafted cues, such as color-based features \cite{chiu2005lane}, the structure tensor~\cite{loose2009kalman}, the bar filter~\cite{teng2010real}, and ridge features~\cite{lopez2010robust}. This information is usually combined with a Hough transform~\cite{liu2010combining, zhou2010novel} and particle or Kalman filters~\cite{kim2008robust, danescu2009probabilistic, teng2010real} to detect lane markings. Then, post-processing methods are leveraged to filter out misdetections and classify lane points to output the final lane detection results~\cite{hillel2014recent}. However, in general, they are prone to effectiveness issues due to road scene variations, e.g., changing from city scene to highway scene and hard to achieve reasonable accuracy under challenging scenarios without a visual clue. 

Recently, deep learning-based segmentation approaches have dominated the lane detection field with more accurate performance~\cite{gopalan2012learning}. For instance, VPGNet~\cite{lee2017vpgnet} proposes a multi-task network for lane marking detection. To better utilize more visual information of lane markings,
SCNN~\cite{pan2018spatial} applies a novel convolution operation that aggregates diverse dimension information via processing sliced features and then adds them together. In order to accelerate the detection speed, light-weight DNNs have been proposed for real-time applications. For example, self-attention distillation (SAD)~\cite{hou2019learning} adopts an attention distillation mechanism. Besides, other methods such as sequential prediction and clustering are also introduced. In~\cite{li2016deep}, a long short-term memory (LSTM) network is presented to face the lane's long line structure issue. Similarly, Fast-Draw~\cite{philion2019fastdraw} predicts the lane's direction at the pixel-wise level. In~\cite{hsu2018learning}, the problem of lane detection is defined as a binary clustering problem. The method proposed in~\cite{hou2019agnostic} also uses a clustering approach for lane detection. Subsequently, a 3D form of lane detection ~\cite{garnett20193d} is introduced to face the non-flatten ground issue.

% \subsubsection{Localization, combine to SLAM as the global localization}

\subsubsection{Localization and mapping}
Localization and mapping are fundamental to autonomous driving. Localization is responsible for finding ego-position relative to a map~\cite{Slamsurvey2018}. The mapping constructs multi-layer high definition (HD) maps~\cite{JIANG2019305} for path planning. Therefore, the accuracy of localization and mapping affects the feasibility and safety of path planning. Currently, GPS-IMU based localization methods have been widely utilized in navigation software like Google Maps. However, the accuracy required for urban automated driving cannot be fulfilled by GPS-IMU systems \cite{darpa2004carnegie}. 

Currently, systems that use a pre-build HD map are more practical and accurate. There are three main types of HD maps: landmark-based, point cloud-based, and vision-based. Landmarks such as poles, curbs, signs, and road markers can be detected with LiDAR~\cite{Lidarlandmark2014} or camera~\cite{CameraLanmark2016}. Landmark searching consumes less computation than the point cloud-based approach but fails in scenarios where landmarks are insufficient. The point cloud contains detailed information about the environment with thousands of points from LiDAR~\cite{Lidarpointclouds2019} or camera~\cite{RGBDpointclouds2016}. Iterative closest point (ICP)~\cite{ICP} and normal distributions transform (NDT)~\cite{NDT} are two algorithms used in point cloud-based HD map generation. They utilize numerical optimization algorithms to calculate the best match. ICP iteratively selects the closest point to calculate the best match.
%between the new scan point cloud and the map to calculate the best match with the minimum distance error. This procedure often takes several iterations until convergence.
On the other side, NDT represents the map as a combination of the normal distribution, then uses the maximum likelihood estimation equation to search match. NDT's computation complexity is less than ICP~\cite{NDTcomparision2009}, but it is not as robust as ICP. Vision-based HD maps are another direction recently becoming more and more popular. The computational overhead limits its application in real systems. Several methods for matching maps with the 2D camera as well as matching 2D image to the 3D image are proposed for mapping~\cite{2Dto3Dmatching2014, 2Dto3Dmatching2020,Depthimagematch}.

In contrast, SLAM \cite{SLAMsurvery2017} is proposed to build the map and localize the vehicle simultaneously. 
% Although it’s less efficient in localization than matching in a pre-built map due to its high computational requirement, SLAM is feasible in the environments where GPS stops functioning. No pre-built map is needed. 
SLAM can be divided into LiDAR-based SLAM and camera-based SLAM. Among LiDAR-based SLAM algorithms, LOAM \cite{Loam} can be finished in real-time. IMLS-SLAM \cite{Imls} focuses on reducing accumulated drift by utilizing a scan-to-model matching method. Cartographer~\cite{Cartographer}, a SLAM package from Google, improves performance by using sub-map and loop closure while supporting both 2D and 3D LiDAR. Compared with LiDAR-based SLAM, camera-based SLAM approaches use frame-to-frame matching. There are two types of matching methods: feature-based and direct matching. Feature-based methods \cite{Orb2vslam, Openvslam, Provslam} extract features and track them to calculate the motion of the camera. Since features are sparse in the image, feature-based methods are also called sparse visual SLAM. Direct matching \cite{Dtamvslam, Rgbdvslam, Rtabvslam} is called dense visual SLAM, which adopts original information for matching that is dense in the image, such as color and depth from an RGB-D camera. 
The inherent properties of feature-based methods lead to its faster speed but tend to fail in texture-less environments as well. The dense SLAM solves the issues of the sparse SLAM with higher computation complexity. For situations that lack computation resources, semiDense \cite{Lsdvslam, Evovslam} SLAM methods that only use direct methods are proposed. Besides the above methods, deep learning methods are also utilized in solving feature extraction \cite{Popupvslam}, motion estimation \cite{Vsovslam}, and long-term localization \cite{Xviewvslam}.

% \subsubsection{Prediction}

\subsubsection{Prediction and planning}

The prediction module evaluates the driving behaviors of the surrounding vehicles and pedestrians for risk assessment~\cite{ADsurvey2020}. Hidden Markov model (HMM) has been used to predict the target vehicle's future behavior and detect unsafe lane change events~\cite{geng2017scenario,yamazaki2016integrating}.

Planning means finding feasible routes on the map from origin to destination. GPS navigation systems are known as global planners~\cite{Globalnavigation2015} to plan a feasible global route, but it does not guarantee safety. In this context, the local planner is developed ~\cite{Localnavigation2016}, which can be divided into three groups: (1) Graph-based planners that give the best path to the destination. (2) Sampling-based planners which randomly scan the environments and only find a feasible path. (3) Interpolating curve planners that are proposed to smooth the path. A*~\cite{Astar} is a heuristic implementation of Dijkstra that always preferentially searches the path from the origin to the destination (without considering the vehicle's motion control), which causes the planning generated by A* to not always be executed by the vehicle. To remedy this problem, hybrid A*~\cite{HybridAstar} generates a drivable curve between each node instead of a jerky line. Sampling-based planners~\cite{Samplingplanner} randomly select nodes for search in the graph, reducing the searching time. Among them, Rapidly-exploring Random Tree (RRT)~\cite{RRT} is the most commonly used method for automated vehicles. As an extension of RRT, RRT*~\cite{RRTstar1,RRTstar2} tries to search the optimal paths satisfying real-time constraints. How to balance the sampling size and computation efficiency is a big challenge for sampling-based planners. Graph-based planners and sampling-based planners can achieve optimal or sub-optimal with jerky paths that can be smoothed with interpolating curve planners. 
% \liangkai{Not clear: These planners sample a set of way-points from paths and fit a given description of the road by considering feasibility, comfort and vehicle dynamics~\cite{Planningsurvey}.} \ren{These planners 
% don't create paths but sample sets of way-points from the trajectories obtained from other planners, then fit sampled points into Specified curves, such as Bezier curves and Polynomial curves. During fitting, the description of roads can be set as constrain to make the generated curves feasibility and comfort.}

\subsubsection{Vehicle control}
Vehicle control connects autonomous driving computing systems and the drive-by-wire system. It adjusts the steering angle and maintains the desired speed to follow the planning module's trajectories. Typically, vehicle control is accomplished by using two controllers: lateral controller and longitudinal controller. Controllers must handle rough and curvy roads, and quickly varying types, such as gravel, loose sand, and mud puddles \cite{StanleylateralControl}, which are not considered by vehicle planners. The output commands are calculated from the vehicle state and the trajectory by control law. There are various control laws, such as fuzzy control \cite{FuzzyControl, FuzzyControlSurvey}, PID control \cite{PIDcontrol, PIDcontrol2}, Stanley control \cite{StanleylateralControl} and Model predictive control (MPC) \cite{MPCcontrol1, MPCcontrol2, MPCcontrol3}. PID control creates outputs based on proportional, integral, and derivative teams of inputs. Fuzzy control accepts continuous values between 0 and 1, instead of either 1 or 0, as inputs continuously respond. Stanley control is utilized to follow the reference path by minimizing the heading angle and cross-track error using a nonlinear control law. MPC performs a finite horizon optimization to identify the control command. Since it can handle various constraints and use past and current errors to predict more accurate solutions, MPC has been used to solve hard control problems like following overtaking trajectories \cite{MPCovertaking}. Controllers derive control laws depending on the vehicle model. Kinematic bicycle models and dynamic bicycle models are most commonly used. In \cite{VehicleModelComparision}, a comparison is present to determine which of these two models is more suitable for MPC in forecast error and computational overhead.

% Kinematic bicycle models and dynamic bicycle models are most commonly used. To learn which model is more suitable for designing MPC, research \cite{VehicleModelComparision} compared in two sides, forecast error and computational overhead. The comparison results showed that the kinematic model performs better forecast than dynamic model \liangkai{Although MPC is computationally expensive and is hard to design the optimization function, several works have been proposed to solve hard control problems, for instance, following overtaking trajectories \cite{MPCovertaking}.} Controllers derive control law depending on the vehicle model. Kinematic bicycle models and dynamic bicycle models are most commonly used. \liangkai{This paper \cite{VehicleModelComparision}, by comparison in forecast error and computational overhead, discussed which of these two models is more suitable for designing MPC.}

\begin{table*}
\centering
\setlength{\extrarowheight}{0pt}
\addtolength{\extrarowheight}{\aboverulesep}
\addtolength{\extrarowheight}{\belowrulesep}
\setlength{\aboverulesep}{0pt}
\setlength{\belowrulesep}{0pt}
\caption{The comparison of different computing hardware for autonomous driving.}
\label{tab: hardware-comparison}
\begin{tabular}{ccccc} 
\toprule
\textbf{Boards}  & \textbf{Architecture}  & \textbf{Performance}  & \textbf{Power Consumption}  & \textbf{Cost\footnotemark[1]}  \\ 
\hline
NVIDIA DRIVE PX2 & GPU & 30 TOPS & 60W & \$15,000 \\
\rowcolor[rgb]{0.839,0.839,0.839} NVIDIA DRIVE AGX & GPU & 320 TOPS & 300W & \$30,000 \\
Texas Instruments TDA3x & DSP & - & 30mW in 30fps & \$549 \\
\rowcolor[rgb]{0.839,0.839,0.839} Zynq UltraScale+ MPSoC ZCU104 & FPGA & 14 images/sec/Watt & - & \$1,295 \\
Mobileye EyeQ5 & ASIC & 24 TOPS & 10W & \$750 \\
\rowcolor[rgb]{0.839,0.839,0.839} Google TPU v3 & ASIC & 420 TFLOPS & 40W & \$8 per hour \\
\bottomrule
\end{tabular}
\end{table*}
\footnotetext[1]{The cost of each unit is based on the price listed on their web site when the product is released to the market.}

\subsection{Computation Hardware}

To support real-time data processing from various sensors, powerful computing hardware is essential to autonomous vehicles' safety. Currently, plenty of computing hardware with different designs show up on the automobile and computing market. In this section, we will show several representative designs based on Graphic Processor Unit (GPU), Digital Signal Processor (DSP), Field Programmable Gate Arrays (FPGA), and Application-Specific Integrated Circuit (ASIC). The comparisons of GPU, DSP, FPGA, and ASIC in terms of architecture, performance, power consumption, and cost are shown in Table~\ref{tab: hardware-comparison}.

NVIDIA DRIVE AGX is the newest solution from NVIDIA unveiled at CES 2018~\cite{nvidia-xavier}. NVIDIA DRIVE AGX is the world’s most powerful System-on-Chip (SoC), and it is ten times more powerful than the NVIDIA Drive PX2 platform. Each DRIVE AGX consists of two Xavier cores. Each Xavier has a custom 8-core CPU and a 512-core Volta GPU. DRIVE AGX is capable of 320 trillion operations per second (TOPS) of processing performance.

% Altera’s Cyclone V SoC is one FPGA-based autonomous driving solution which has been used in Audi products. Altera’s FPGAs are optimized for sensor fusion, combining data from multiple sensors in the vehicle for highly reliable object detection~\cite{FPGA1}. Similarly, Zynq UltraScale MPSoC is also designed for autonomous driving tasks~\cite{FPGA2}. When running Convolution Neural Network tasks, it achieves 14 images/sec/Watt, which outperforms the Tesla K40 GPU (4 images/sec/Watt). Also, for object tracking tasks, it reaches 60 fps in a live 1080p video stream.

Zynq UltraScale+ MPSoC ZCU104 is an automotive-grade product from Xilinx~\cite{xilinx-news}. It is an FPGA-based device designed for autonomous driving. It includes 64-bit quad-core ARM® Cortex™-A53 and dual-core ARM Cortex-R5. This scalable solution claims to deliver the right performance/watt with safety and security~\cite{xilinx-AD}. When running CNN tasks, it achieves 14 images/sec/watt, which outperforms the Tesla K40 GPU (4 images/sec/watt). Also, for object tracking tasks, it reaches 60 fps in a live 1080p video stream.

Texas Instruments’ TDA provides a DSP-based solution for autonomous driving. A TDA3x SoC consists of two C66x Floating-Point VLIW DSP cores with vision AccelerationPac. Furthermore, each TDA3x SoC has dual Arm Cortex-M4 image processors. The vision accelerator is designed to accelerate the process functions on images. Compared with an ARM Cortex-15 CPU, TDA3x SoC provides an eight-fold acceleration on computer vision tasks with less power consumption~\cite{DSP1}. 
% Similarly, CEVA XM4 is another DSP-based autonomous driving computing solution. It is designed for computer vision tasks on video streams. The main benefit for using CEVA-XM4 is energy-efficiency, which requires less than 30mW for a 1080p video at 30 frames per second~\cite{DSP2}.

MobileEye EyeQ5 is the leading ASIC-based solution to support fully-autonomous (Level 5) vehicles~\cite{mobileeye-Q5}. EyeQ5 is designed based on 7nm-FinFET semiconductor technology, and it provides 24Tops computation capability with 10 watts' power budget. TPU is Google's AI accelerator ASIC mainly for neural network and machine learning~\cite{google-tpu}. TPU v3 is the newest release, which provides 420 TFLOPS computation for a single board. 

\subsection{Storage}
%HydraSpace
The data captured by an autonomous vehicle is proliferating, typically generating between 20TB and 40TB per day, per vehicle~\cite{AutoTech}. The data includes cameras (20 to 40MB), as well as sonar (10 to 100KB), radar (10 to 100KB), and LiDAR (10 to 70MB)~\cite{IOT,neon}. Storing data securely and efficiently can  accelerate overall system performance. Take object detection for example: the history data could contribute to the improvement of detection precision using machine learning algorithms. Map generation can also benefit from the stored data in updating traffic and road conditions appropriately. 
Additionally, the sensor data can be utilized to ensure public safety and predict and prevent crime. The biggest challenge is to ensure that sensors collect the right data, and it is processed immediately, stored securely, and transferred to other technologies in the chain, such as Road-Side Unit (RSU), cloud data center, and even third-party users~\cite{zhang2018openvdap}. More importantly, creating hierarchical storage and workflow that enables smooth data accessing and computing is still an open question for the future development of autonomous vehicles. 

In~\cite{wang-hydraspace}, a computational storage system called HydraSpace is proposed to tackle the storage issue for autonomous driving vehicles. HydraSpace is designed with multi-layered storage architecture and practical compression algorithms to manage the sensor pipe data. OpenVDAP is a full-stack edge-based data analytic platform for connected and autonomous vehicles (CAVs)~\cite{zhang2018openvdap}. It envisions for the future four types of CAVs applications, including autonomous driving, in-vehicle infotainment, real-time diagnostics, and third-party applications like traffic information collector and SafeShareRide~\cite{liu2018safeshareride}. The hierarchical design of the storage system called driving data integrator (DDI) is proposed in OpenVDAP to provide sensor-aware and application-aware data storage and processing~\cite{zhang2018openvdap}.

\subsection{Real-Time Operating Systems}

According to the automation level definitions from the SAE~\cite{Self-driving-car}, the automation of vehicles increases from level 2 to level 5, and the level 5 requires full automation of the vehicle, which means the vehicle can drive under any environment without the help from the human. To make the vehicle run in a safe mode, how to precept the environment and make decisions in real-time becomes a big challenge. That is why real-time operating systems become a hot topic in the design and implementation of autonomous driving systems.

% \subsubsection{Real-time operating system}
%QNX, VxWorks, RT-linux

% Operating system is the system software which manages the computing resources and enables the execution of applications on the system~\cite{silberschatz2018operating}. A Real-Time Operating System (RTOS) is a special operating system which is proposed to guarantee the deadline of safety-critical tasks such as industrial, automotive, aviation, military etc \cite{jensen1985time}. As a safety critical system, autonomous driving is recognized as a representative scenario of the RTOS. 

% Compared with RTOS which is proposed to provide a general system to handle all the tasks, run time system is proposed to leverage the RTOS to accelerate the execution of tasks on the hardware/software. For example, deep learning based algorithms are very popular in the computer vision-based perception algorithms and heterogeneous hardware is applied to accelerate their performance. However, the general RTOS is not customized for the hardware/architecture and algorithms. Therefore, we need a runtime system to optimize the execution of deep learning algorithms on the heterogeneous hardware. Usually runtime system can be a complimentary part to the RTOS.

RTOS is widely used in the embedded system of ECUs to control the vehicle's throttle, brake, etc. \textit{QNX} and \textit{VxWorks} are two representative commercialized RTOS widely used in the automotive industry. The \textit{QNX} kernel contains only CPU scheduling, inter-process communication, interrupt redirection, and timers. Everything else runs as a user process, including a unique process known as ``proc,'' which performs process creation and memory management by operating in conjunction with the microkernel~\cite{hildebrand1992architectural}. \textit{VxWorks} is designed for embedded systems requiring real-time, deterministic performance and, in many cases, safety and security certification~\cite{VxWorks}. \textit{VxWorks} supports multiple architectures, including Intel, POWER, and ARM. \textit{VxWorks} also uses real-time kernels for mission-critical applications subject to real-time constraints, which guarantees a response within pre-defined time constraints.

\textit{RTLinux} is a microkernel-based operating system that supports hard real-time~\cite{yodaiken1999rtlinux}. The scheduler of \textit{RTLinux} allows full preemption. Compared with using a low-preempt patch in \textit{Linux}, \textit{RTLinux} allows preemption for the whole Linux system. \textit{RTLinux} makes it possible to run real-time critical tasks and interprets them together with the \textit{Linux}~\cite{sato2000real}.

\textit{NVIDIA DRIVE OS} is a foundational software stack from NVIDIA, which consists of an embedded RTOS, hypervisor, NVIDIA CUDA libraries, NVIDIA Tensor RT, etc. that is needed for the acceleration of machine learning algorithms~\cite{nvidia-driveos}. 

\subsection{Middleware Systems}

% ROS, ROS2, Cyber, etc.

Robotic systems, such as autonomous vehicle systems, often involve multiple services, with many dependencies. Middleware is required to facilitate communications between different autonomous driving services.

Most existing autonomous driving solutions utilize the \textit{ROS} \cite{quigley2009ros}. Specifically, \textit{ROS} is a communication middleware that facilitates communications between different modules of an autonomous vehicle system. \textit{ROS} supports four communication methods: topic, service, action, and parameter. \textit{ROS2} is a promising type of middleware developed to make communications more efficient, reliable, and secure~\cite{ros2}. However, most of the packages and tools for sensor data process are still currently based on \textit{ROS}. 

The \textit{Autoware Foundation} is a non-profit organization supporting open-source projects enabling self-driving mobility~\cite{autoware-foundation}. \textit{Autoware.AI} is developed based on \textit{ROS}, and it is the world's first "all-in-one" open-source software for autonomous driving technology. \textit{Apollo Cyber}~\cite{apollocyber.org} is another open-source middleware developed by Baidu. \textit{Apollo} aims to accelerate the development, testing, and deployment of autonomous vehicles. \textit{Apollo Cyber} is a high-performance runtime framework that is greatly optimized for high concurrency, low latency, and high throughput in autonomous driving.

In traditional automobile society, the runtime environment layer in Automotive Open System Architecture(\textit{AutoSAR})~\cite{autosar.org} can be seen as middleware. Many companies develop their middleware to support \textit{AutoSAR}. However, there are few independent open-source middlewares nowadays because it is a commercial vehicle company's core technology. Auto companies prefer to provide middleware as a component of a complete set of autonomous driving solutions.

\subsection{Vehicular Communication}

In addition to obtaining information from the on-board sensors, the recent proliferation in communication mechanisms, e.g., DSRC, C-V2X, and 5G, has enabled autonomous driving vehicles to obtain information from other vehicles, infrastructures like traffic lights and RSU as well as pedestrians. 

\subsubsection{LTE/4G/5G}

Long-Term Evolution (LTE) is a transitional product in the transition from 3G to 4G \cite{liu2020equinox}, which provides downlink peak rates of 300 Mbit/s, uplink peak rates of 75 Mbit/s. The fourth-generation communications (4G) comply with 1 Gbit/s for stationary reception and 100 Mbit/s for mobile. As the next-generation mobile communication, U.S. users that experienced the fastest average 5G download speed reached 494.7 Mbps on Verizon, 17.7 times faster than that of 4G. And from Verizon's early report, the latency of 5G is less than 30 ms, 23 ms faster than average 4G metrics. However, we cannot deny that 5G still has the following challenges: complex system, high costs, and poor obstacle avoidance capabilities.

\subsubsection{DSRC}

DSRC \cite{kenney2011dedicated} is a type of V2X communication protocol, which is specially designed for connected vehicles. DSRC is based on the IEEE 802.11p standard, and its working frequency is 5.9GHz. Fifteen message types are defined in the SAE J2735 standard~\cite{standard2009dedicated}, which covers information like the vehicle's position, map information, emergence warning, etc.~\cite{kenney2011dedicated}. Limited by the available bandwidth, DSRC messages have small size and low frequency. However, DSRC provides reliable communication, even when the vehicle is driving 120 miles per hour.

% \liangkai{As is based on ad hoc network, the throughput performance of DSRC will significantly the number of the nodes increases because of the traffic congestion \cite{xu2017dsrc}.}

\subsubsection{C-V2X}
C-V2X combines the traditional V2X network with the cellular network, which delivers mature network assistance and commercial services of 4G/5G into autonomous driving. Like DSRC, the working frequency of C-V2X is also the primary common spectrum, 5.9 GHz~\cite{5g2016case}. Different from the CSMA-CA in DSRC, C-V2X has no contention overheads by using semi-persistent transmission with relative energy-based selection. Besides, the performance of C-V2X can be seamlessly improved with the upgrade of the cellular network. Generally, C-V2X is more suitable for V2X scenarios where cellular networks are widely deployed.

\subsection{Security and Privacy}
%sensor security,
%computing device security,
%data security,
%runtime security,
%privacy
With the increasing degree of vehicle electronification and the reliance on a wide variety of technologies, such as sensing and machine learning, the security of AVs has risen from the hardware damage of traditional vehicles to comprehensive security with multi-domain knowledge. Here, we introduce several security problems strongly associated with AVs with the current attacking methods and standard coping methods. In addition to the security and privacy issues mentioned as follows, AVs systems should also take care of many other security issues in other domains, such as patching vulnerabilities of hardware or software systems and detecting intrusions \cite{intrusiondetection}. 
\subsubsection{Sensing security}

As the eye of autonomous vehicles, the security of sensors is nearly essential. Typically, jamming attacks and spoofing attacks are two primary attacks for various sensors \cite{attackPIEEE, sensorattacks}. For example, the spoofing attack generates an interference signal, resulting in a fake obstacle captured by the vehicle \cite{yan2016sensorattack}. Besides, GPS also encounters spoofed attacks~\cite{gpsattack}. Therefore, protection mechanisms are expected for sensor security. Randomized signals and redundant sensors are usually used by these signal-reflection sensors \cite{shin2017def,petit2015remote}, including LiDAR and radar. The GPS can check signal characteristics \cite{dlr2013gpsdef} and authenticate data sources \cite{gpsencryption} to prevent attacks. Also, sensing data fusion is an effective mechanism. 

\subsubsection{Communication security}

Communication security includes two aspects: internal communication and outside communication. Currently, internal communication like CAN, LIN, and FlexRay, has faced severe security threats \cite{canbussecurity,LINbussecurity,FlexRaysecurity}. The cryptography is frequently-used technology to keep the transmitted data confidential, integrated, and authenticated \cite{stinson2018cryptography}. However, the usage of cryptography is limited by the high computational cost for these resource-constrained ECUs. Therefore, another attempt is to use the gateway to prevent unallowed access \cite{Kim2015gateway}. The outside communication has been studied in VANETs with V2V, V2R, and V2X communications \cite{communicationsecurity,Qu2015vanets,Ali2019vanets}. Cryptography is the primary tool. A trusted key distribution and management is built in most approaches, and vehicles use assigned keys to authenticate vehicles and data. 

\subsubsection{Data security}

Data security refers to preventing data leakage from the perspectives of transmission and storage. The former has been discussed in communication security, where various cryptography approaches are proposed to protect data in different scenarios \cite{Zhong2019msgauth,Garg2019securityframework}. The cryptography is also a significant technology of securing data storage, such as an encrypted database \cite{cryptdb} and file system \cite{encryptedfilesystem}. Besides, access control technology \cite{Sandhu1994accesscontrol} protects stored data from another view, widely-used in modern operating systems. An access control framework \cite{zhang2020ac4av} has been proposed for AVs to protect in-vehicle data in real-time data and historical data, with different access control models. 

\subsubsection{Control security}

With vehicles' electronification, users could open the door through an electronic key and control their vehicles through an application or voice. However, this also leads to new attack surfaces with various attack methods, such as jamming attacks, replay attacks, relay attacks, etc.~\cite{attackPIEEE}. For example, attackers could capture the communication between key and door and replay it to open the door~\cite{Kamkar2015def}. Also, for those voice control supported vehicles, the attackers could successfully control the vehicle by using voices that humans cannot hear~\cite{zhang2017DolphinAttack}. Parts of these attacks could be classified into sensing security, communication security, or data security, which can be addressed by corresponding protection mechanisms.

\subsubsection{Privacy}
Autonomous vehicles heavily rely on the data of the surrounding environment, which typically contains user privacy. For example, by recognizing buildings in cameras, attackers can learn the vehicle location \cite{XiongCameraLocation}. Or an attacker can obtain the location directly from GPS data. Thus, the most straightforward but the most difficult solution is to prevent data from being obtained by an attacker, such as access control \cite{Sandhu1994accesscontrol,zhang2020ac4av} and data encryption \cite{attackPIEEE}. However, autonomous vehicles will inevitably utilize location-based services. Except for the leak of current location, attackers could learn the home address from the vehicle trajectory \cite{dataprivacyinCAV}. Thus, data desensitization is necessary to protect privacy, including anonymization and differential privacy \cite{Martinelli2020driveridentify}.
%An attacker can learn user privacy from analyzing user data. For example, by analyzing vehicle control data, the driver identity can be recognized \cite{Martinelli2020driveridentify}. Thus, the most straightforward but hard protection is to prevent data from being obtained by an attacker, such as access control and data encryption. Another way is data desensitization, including anonymization and differential privacy \cite{dataprivacyinCAV}. 

\section{Challenges and Discussions}
\label{challenges}

% \textcolor{red}{\textbf{each subsection half page (one column)}}
% \liangkai{need add a connection paragraphs here overviewing}

From the review of the current key technologies of the computing system for autonomous driving, we find that there are still many challenges and open issues for the research and development of L4 or L5 autonomous driving vehicles. In this section, we summarize twelve remaining challenges and discuss the challenges with our visions for autonomous driving.

\subsection{Artificial intelligence for AVs}

Most of the AV's services (e.g., environmental perception and path planning) are carried out by artificial-intelligence-based approaches. As the focus of the automotive industry gradually shifts to series production, the main challenge is how to apply machine learning algorithms to mass-produced AVs for real-world applications. Here, we list three main challenges in artificial intelligence (AI) for AVs. 

\subsubsection{Standardization of safety issue} one of the main challenges is that machine learning algorithms are unstable in terms of performance. For example, even a small change to camera images (such as cropping and variations in lighting conditions) may cause the ADAS system to fail in object detection and segmentation \cite{pan2018SCNN, ghafoorian2018gan, xu2017end}. However, the automotive safety standard of ISO 26262 \cite{hommes2012review} was defined without taking deep learning into consideration because the ISO 26262 was published before the boom of AI, leading to the absence of proper ways to standardize the safety issue when incorporating AI for AVs \cite{8452728}.

\subsubsection{Infeasibility of scalable training} to achieve high performance, machine learning models used on AVs need to be trained on representative datasets under all application scenarios, which bring challenges in training time-sensitive models based on Petabytes of data. In this case, collaborative training \cite{lu2019collaborative}, model compression technologies \cite{courbariaux2015training, han2015learning, denton2014exploiting, denil2013predicting, sau2016deep, luo2016face}, and lightweight machine learning algorithms \cite{howard2017mobilenets, chollet2017xception, iandola2016squeezenet} were proposed in recent years. Besides, getting accurate annotations of every pedestrian, vehicle, lane, and other objects are necessary for the model training using supervised learning approaches, which becomes a significant bottleneck \cite{yu2018bdd100k}.

\subsubsection{Infeasibility of complete testing}  it is infeasible to test machine learning models used on AVs thoroughly. One reason is that machine learning learns from large amounts of data and stores the model in a complex set of weighted feature combinations, which is not intuitive or difficult to conduct thorough testing \cite{koopman2016challenges}. In addition, previous work pointed out that, to verify the catastrophic failure rate, around $10^9$ hours (billion hours) of vehicle operation test should be carried out \cite{butler1991infeasibility} and the test needs be repeated many times to achieve statistical significance \cite{8452728}.

\subsection{Multi-sensors Data Synchronization}
% reliability

Data on the autonomous driving vehicle has various sources: its sensors, other vehicle sensors, RSU, and even social media. One big challenge to handle a variety of data sources is how to synchronize them. 

For example, a camera usually produces 30-60 frames per second, while LiDAR's point cloud data frequency is 10HZ. For applications like 3D object detection, which requires camera frames and point cloud at the same time, should the storage system do synchronization beforehand or let the application developer do it? This issue becomes more challenging, considering that the timestamp's accuracy from different sensors falls into different granularities. For example, considering the vehicles that use network time protocol (NTP) for time synchronization, the timestamp difference can be as long as 100ms~\cite{mills1992rfc1305,GPS-accuracy}. For some sensors with a built-in GNSS antenna, the time accuracy goes to the nanosecond level. In contrast, other sensors get a timestamp from the host machine's system time when accuracy is at the millisecond level. Since the accuracy of time synchronization is expected to affect the vehicle control's safety, handling the sensor data with different frequency and timestamp accuracy is still an open question.

\subsection{Failure Detection and Diagnostics}

% sensor failure, sensor data failure (sensor is fine but the sensor data cannot be used, like something block the lens of camera), processing algorithm failure

% Today's autonomous vehicles (AVs) are equipped with multiple sensors to achieve safe and reliable navigation and precise perception of the environment. These sensors include cameras, LiDARs, radars, and~GPS \cite{yan2016can}. 

Today's AVs are equipped with multiple sensors, including LiDARs, radars, and~GPS \cite{yan2016can}. Although we can take advantage of these sensors in terms of providing a robust and complete description of the surrounding area, some open problems related to the failure detection are waiting to be solved. Here, we list and discuss four failure detection challenges: (1) \textit{Definition of sensor failure:} there is no standard, agreed-upon universal definition or standards to define the scenario of sensor failures~\cite{sabaliauskaite2018integrating}. However, we must propose and categorize the standard of sensor failures to support failure detection by applying proper methods. (2) \textit{Sensor failure:} more importantly, there is no comprehensive and reliable study on sensor failure detection, which is extremely dangerous since most of the self-driving applications are relying on the data produced by these sensors \cite{orrick1994failure}. If some sensors encountered a failure, collisions and environmental catastrophes might happen. (3) \textit{Sensor data failure:} in the real application scenario, even when the sensors themselves are working correctly, the generated data may still not reflect the actual scenario and report the wrong information to people \cite{tian2018deeptest}. For instance, the camera is blocked by unknown objects such as leaves or mud, or the radar deviates from its original fixed position due to wind force. In this context, sensor data failure detection is very challenging, (4) \textit{Algorithm failure:} In challenging scenarios with severe occlusion and extreme lighting conditions, such as night, rainy days, and snowy days, deploying and executing state-of-the-art algorithms cannot guarantee output the ideal results \cite{tang2020performance}. For example, lane markings usually fail to be detected at night by algorithms that find it difficult to explicitly utilize prior information like rigidity and smoothness of lanes \cite{tamai1996ego}. However, humans can easily infer their positions and fill in the occluded part of the context. Therefore, how to develop advanced algorithms to further improve detection accuracy is still a big challenge.

For a complex system with rich sensors and hardware devices, failures could happen everywhere. How to tackle the failure and diagnosing the issue becomes a big issue. One example is the diagnose of lane controller systems from Google~\cite{lee2015system}. The idea is to determine the root cause of malfunctions based on comparing the actual steering corrections applied to those predicted by the virtual dynamics module.

\subsection{How to Deal with Normal-Abnormal?}
% rain/winter case, workzone, emergency scenarios, etc.

% how to detect the normal-abnormal cases and how the algorithms / systems handle it (use example to explain).
Normal-abnormal represents normal scenarios in daily life that are, abnormal in the autonomous driving dataset. Typically, there are three cases of normal-abnormal: adverse weather, emergency maneuvers, and work zones.

\subsubsection{Adverse weather}

One of the most critical issues in the development of AVs is the poor performance under adverse weather conditions, such as rain, snow, fog, and hail, because the equipped sensors (e.g., LiDAR, radar, camera, and GPS) might be significantly affected by the extreme weather. The work of ~\cite{8666747}  characterized the effect of rainfall on millimeter-wave (mm-wave) radar and proved that under heavy rainfall conditions, the detection range of millimeter-wave radar can be reduced by as much as 45\%. Filgueira et al.~\cite{filgueira2017quantifying} pointed out that as the rain intensity increases, the detected LiDAR intensity will attenuate. At the same time, Bernardin et al. \cite{bernardin2014measuring} proposed a methodology to quantitatively estimate the loss of visual performance due to rainfall. Most importantly, experimental results show that, compared to training in narrow cases and scenarios, using various data sets to train object detection networks may not necessarily improve the performance of these networks.~\cite{hnewa2020object}. However, there is currently no research to provide a systematic and unified method to reduce the impact of weather on various sensors used in AVs. Therefore, there is an urgent need for novel deep learning networks that have sufficient capabilities to cope with safe autonomous driving under severe weather conditions.

\subsubsection{Emergency maneuvers}
In emergency situations, such as a road collapse, braking failure, a tire blowout, or suddenly seeing a previously ``invisible'' pedestrian, the maneuvering of the AVs may need to reach its operating limit to avoid collisions. 
However, these collision avoidance actions usually conflict with stabilization actions aimed at preventing the vehicle from losing control, and in the end, they may cause collision accidents. In this context, some research has been done to guarantee safe driving for AVs in emergent situations. For example, Hilgert et al. proposed a path planning method for emergency maneuvers based on elastic bands \cite{1225546}. ~\cite {7585053} is proposed to determine the minimum distance at which obstacles cannot be avoided at a given speed. Guo et al. \cite{7457666} discussed dynamic control design for automated driving, with particular emphasis on coordinated steering and braking control in emergency avoidance. Nevertheless, how an autonomous vehicle safely responds to different classes of emergencies with on-board sensors is still an open problem.

% In these scenarios, vehicle stabilization becomes important to ensure that the vehicle does not lose control. However, stabilization actions may conflict with those necessary for collision avoidance, which potentially leading to a collision.  is extremely dangerous for drivers on the road. Therefore, how an autonomous vehicle must safely respond to different classes of emergencies using vision and other on-board sensors is also an open problem.

% how to guarantee the safe driving for AVs in emergency situation

\subsubsection{Work zone}

Work zone recognition is another challenge for an autonomous driving system to overcome. For most drivers, the work zone means congestion and delay of the driving plan. Many projects have been launched to reduce and eliminate work zone injuries and deaths for construction workers and motorists. "Workzonesafety.org" summarizes recent years of work zone crashes and supplies training programs to increase public awareness of the importance of work-zone safety. 
% Automated vehicles are attractive in aspects reducing crashes and increasing road safety, only few researchers are working on work zone related autonomous driving system. 
Seo~\cite{Seo2014} proposed a machine learning-based method to improve the recognition of work zone signs. Developers from Kratos Defense \& Security Solutions~\cite{ATMA2019} present an autonomous truck which safely passes a work zone. Their system relied on V2V communications to connect the self-driving vehicle with a leader vehicle. The self-driving vehicle accepted navigation data from the leader vehicle to travel along its route while keeping a pre-defined distance. 
Until now, the work zone is still a threat to drivers and workers' safety but has not attracted too much attention to autonomous driving researchers. There are still significant gaps in this research field, waiting for researchers to explore and tackle critical problems.

\subsection{Cyberattack Protection}

Attacks and defenses are always opposites, and absolute security does not exist. The emerging CAVs face many security challenges, such as reply attacks to simulate a vehicle's electronic key and spoof attacks to make vehicle detour \cite{Kamkar2015def,attackPIEEE}. With the integration of new sensors, devices, technologies, infrastructures, and applications, the attack surface of CAVs is further expanded. 

Many attacks focus on one part of the CAVs system and could be protected by the method of fusing several other views. For example, a cheated roadblock detected by radars could be corrected by camera data. Thus, how to build such a system to protect CAVs, systematically, is the first challenge for the CAVs system. The protection system is expected to detect potential attacks, evaluate the system security status, and recover from attacks.

Besides, some novel attack methods should be attended. Recently, some attacks have been proposed to trick these algorithms \cite{algorAttack}. For example, a photo instead of a human to pass the face recognition or a note-sized photo posted on the forehead makes machine learning algorithms fail to detect faces \cite{komkov2019advhat}. Thus, how to defend the attacks on machine learning algorithms is a challenge for CAVs systems.

Furthermore, some new technologies could be used to enhance the security of the CAVs system. 
With the development of quantum computing technology, the existing cryptography standards cannot ensure protected data, communication, and systems. Thus, designing post-quantum cryptography \cite{postquantumcrypt} and architecture is a promising topic for CAVs and infrastructure in ITS. 
% In addition, various applications will be installed on the vehicle, in-vehicle data security and privacy are also important.

Also, we noticed that the hardware-assistant trusted execution environment \cite{Ning2017tee} could improve the system security, which provides an isolated and trusted execution environment (TEE) for applications. However, it has limited physical memory size, and execution performance will drop sharply as the total memory usage increases. Therefore, how to split the system components and make critical parts in the TEE with high security is still a challenge in design and implementation.

\subsection{Vehicle Operating System}

The vehicle operating system is expected to abstract the hardware resources for higher layer middleware and autonomous driving applications. In the vehicle operating system development, one of the biggest challenges is the compatibility with the vehicle's embedded system. Take Autoware as an example: although it is a full-stack solution for the vehicle operating system that provides a rich set of self-driving modules composed of sensing, computing, and actuation capabilities, the usage of it is still limited to several commercial vehicles with a small set of supportable sensors~\cite{kato2018autoware}. On a modern automobile, as many as 70 ECUs are installed for various subsystems, and they are communicated via CAN bus. For the sake of system security and commercial interests, most of the vehicles' CAN protocol is not open-sourced, which is the main obstacle for developing a unified vehicle operating system. 

AUTOSAR is a standardization initiative of leading automotive manufacturers and suppliers founded in the autumn of 2003~\cite{autosar.org}. AUTOSAR is promising in narrowing the gap for developing an open-source vehicle operating system. However, most automobile companies are relatively conservative to open-source their vehicle operating systems, restricting the availability of AUTOSAR to the general research and education community. There is still a strong demand for a robust, open-source vehicle operating system for AVs.

\subsection{Energy Consumption}
With rich sensors and powerful computing devices implemented on the vehicle, energy consumption becomes a big issue. Take the NVIDIA Drive PX Pegasus as an example: it consumes 320  INT8  TOPS  of  AI  computational  power  with a 500  watts budget. If we added external devices like sensors, communication antennas, storage, battery, etc., the total energy consumption would be larger than 1000W~\cite{nvidia-xavier}. Besides, if a duplicate system is installed for the autonomous driving applications' reliability, the total power dissipation could go up to almost 2000W.

How to handle such a tremendous amount of power dissipation is not only a problem for the battery management system; it is also a problem for the heat dissipation system. What makes this issue more severe is the size limitation and auto-grid requirements from the vehicle's perspective. How to make the computing system of the autonomous driving vehicle become energy efficient is still an open challenge. E2M tackles this problem by proposing as an energy-efficient middleware for the management and scheduling deep learning applications to save energy for the computing device~\cite{liu2019e2m}. However, according to the profiling results, most of the energy is consumed by vehicles' motors. Energy-efficient autonomous driving requires the co-design in battery cells, energy management systems, and autonomous vehicle computing systems. 

\subsection{Cost}
In the United States, the average cost to build a traditional non-luxury vehicle is roughly \$30,000, and for an AV, the total cost is around \$250,000 \cite{LeVine:2017}. AVs need an abundance of hardware equipment to support their normal functions. Additional hardware equipments required for AVs, include but are not limited to, the communication device, computing equipment, drive-by-wire system, extra power supply, various sensors, cameras, LiDAR, and radar. %The approximate prices of each equipment are shown in Table~\ref{tab:hardwarecost}. 
In addition, to ensure AV's reliability and safety, a backup of these hardware devices may be necessary \cite{Sedgwick:2017}. For example, if the main battery fails, the vehicle should have a backup power source to support computing systems to move the vehicle. 

% \usepackage{multirow}
% \begin{table}[ht]
% \centering
% \caption{Hardware Equipment of Autonomous Vehicles.}
% \label{tab:hardwarecost}
% \scalebox{0.95}{
% \begin{threeparttable}[b]
% \begin{tabular}{|c|c|c|c|}
% \hline
% \multicolumn{2}{|c|}{Hardware Equipment} & \multicolumn{2}{c|}{Unit Price} \\ \hline
% \multicolumn{2}{|c|}{\multirow{2}{*}{Communication Device}} & DSRC & \$17,600 \\ \cline{3-4} 
% \multicolumn{2}{|c|}{} & LTE/5G/C-V2X & \$177 \\ \hline
% \multicolumn{2}{|c|}{Computation Device} & \begin{tabular}[c]{@{}c@{}}Consists of a PC, a hard drive,\\ a graphics card, and other devices\end{tabular} & \$5000 \\ \hline
% \multicolumn{2}{|c|}{\begin{tabular}[c]{@{}c@{}}Vehicle Electronic\\ Control Unit (ECU)\end{tabular}} & Each vehicle has 50 to 100 ECUs & \$320 each \\ \hline
% \multicolumn{2}{|c|}{Power Supply} & \multicolumn{2}{c|}{\$24,300} \\ \hline
% \multirow{4}{*}{\begin{tabular}[c]{@{}c@{}}\\\\\\Sensor\end{tabular}} & Camera & \multicolumn{2}{c|}{\$6000} \\ \cline{2-4} 
%  & LiDAR & \multicolumn{2}{c|}{\begin{tabular}[c]{@{}c@{}}\$8,5000 (installed on the top, \\ 360-degree field of view)\\ + \$8,000 * 4 (4 small ones installed \\ in 4 corners)\end{tabular}} \\ \cline{2-4} 
%  & Radar & \multicolumn{2}{c|}{\$500 each * 6} \\ \cline{2-4} 
%  & Other & \multicolumn{2}{c|}{Eg. inertial measurement unit, \$4,000} \\ \hline
% \end{tabular}
% \begin{tablenotes}
%      \item[1] Data listed as of August 2020.
% \end{tablenotes}
% \end{threeparttable}
% }
% \vspace{-0.6 cm}
% \end{table}

The cost of building an autonomous vehicle is already very high, not to mention the maintenance cost of an AV, e.g., diagnostics and repair. High maintenance costs lead to declining consumer demand and undesirable profitability for the vehicle manufacturers. Companies like Ford and GM have already cut their low-profit production lines to save costs~\cite{Guilford:2018,Luft:2020}. 

Indeed, the cost of computing systems for AVs currently in the research and development stage is very high. However, we hope that with the maturity of the technologies and the emergence of some alternative solutions, the price will ultimately drop to a level that individuals can afford. Take battery packs of electric vehicles (EVs) as an example: when the first mass-market EVs were introduced in 2010, their battery packs were estimated at \$1,000 USD per kilowatt-hour (kWh). However, Tesla’s Model 3 battery pack costs \$190 per kilowatt-hour, and General Motors’ 2017 Chevrolet Bolt battery pack is estimated to cost \$205 per kilowatt-hour. In 6 years, the price per kilowatt-hour has dropped by more than 70\% \cite{Lifespan:2018}. Also, Waymo claims to have successfully reduced the experimental version of high-end LiDAR to approximately \$7,500. 
Besides, Tesla, which uses only radar instead of LiDAR, says its autonomous vehicle equipment is around \$8,000 \cite{LeVine:2017}. In addition to the reduction of hardware costs, we believe that the optimization of computing software in an AV can also help reduce the cost to a great extent.
% According to the “Automated Driving Roadmap” of the European Road Transport Research Advisory Council (ERTRAC) \cite{ERTRAC:2019}, AVs will achieve full automation level (level 5) after 2030, then the era of AVs' profitability might has truly arrived.

% forecast of the European Road Transport Research Advisory Council (ERTRAC) 2019 Roadmap \cite{ERTRAC:2019}, AVs will be after 2030 At some point when fully automated (Level 5) is achieved, maybe the era of AVs' profitability will really come.

\subsection{How to Benefit from Smart Infrastructure?}
% DoT proposal, Dr. Dong's paper
Smart infrastructure combines sensors, computing platforms, and communication devices with the physical traffic infrastructure~\cite{CSIC2016}. It is expected to enable the AVs to achieve more efficient and reliable perception and decision making. Typically, AVs could benefit from smart infrastructure in three aspects: 
(1)~\textit{Service provider} It is struggling for an AV to find a parking space in the parking lot. By deploying sensors like RFID on the smart infrastructure, parking services can be handled quickly~\cite{Pala2007}. As the infrastructure becomes a provider for parking service, it is possible to schedule service requests to achieve the maximum usage. Meanwhile, AVs can reduce the time and computation for searching services. (2)~\textit{Traffic information sharing:} Traffic information is essential to safe driving. Lack of traffic information causes traffic congestion or even accidents. Roadside Units (RSUs) is implemented to provide traffic information to passing vehicles through V2X communications. Besides, RSUs are also used to surveil road situations using various on-board sensors like cameras and LiDARs~\cite{Dweik2017}. The collected data is used for various tasks, including weather warning, map updating, road events detection, and making up blind spots of AVs. (3) \textit{Task offloading:} Various algorithms are running on vehicles for safe driving. Handling all workloads in real-time requires a tremendous amount of computation and power, infeasible on a battery-powered vehicle~\cite{Lin2019}. Therefore, offloading heavy computation workloads to the infrastructure is proposed to accelerate the computation and save energy. However, to perform feasible offloading, the offloading framework must offload computations to the infrastructure while ensuring timing predictability~\cite{Dong2020}. Therefore, how to schedule the order of offloading workloads is still a challenge to benefit from the smart infrastructure.

\subsection{Dealing with Human Drivers}
According to NHTSA data collected from all 50 states and the District of Columbia, 37,461 lives were lost on U.S. roads in 2016, and 94\% of crashes were associated with “a human choice or error”~\cite{releases2016fatal}. Although autonomous driving is proposed to replace human drivers with computers/machines for safety purposes, human driving vehicles will never disappear. How to enable computers/machines in AVs to interact with a human driver becomes a big challenge~\cite{WEF-safedrive}.

Compared with a human driver, machines are generally more suited for tasks like vehicle control and multi-sensor data processing. In contrast, the human driver maintains an advantage in perception and sensing the environment~\cite{schoettle2017sensor}. One of the fundamental reasons is that the machine cannot think like a human. Current machine learning-based approaches cannot handle situations that are not captured in the training dataset. For example, in driving automation from SAE, one of the critical differences between level 2 and level 3/4/5 is whether the vehicle can make decisions like overtaking or lane changing by itself~\cite{liu2019edge}. In some instances, interacting with other human drivers becomes a big challenge because human drivers can make mistakes or violate traffic rules. 

Many works focus on getting a more accurate speed and control predictions of the surrounding vehicles to handle the machine-human interaction~\cite{hubmann2017decision,geng2017scenario}. Deep reinforcement learning shows promising performance in complex scenarios requiring interaction with other vehicles~\cite{li2016deep,sallab2017deep}. However, they are either simulation-based or demonstration in limited scenarios. Another promising direction to tackle machine-human interaction is through V2X communications. Compared with predicting other vehicles' behavior, it is more accurate to communicate safety information~\cite{deng2019cooperative}. 

\subsection{Experimental Platform}
The deployment of autonomous driving algorithms or prototypes requires complex tests and evaluations in a real environment, which makes the experimental platform becomes one of the fundamental parts of conducting research and development. However, building and maintaining an autonomous driving vehicle is enormous: the cost of a real autonomous driving vehicle could attain \$250,000; maintaining the vehicle requires parking, insurance, and auto maintenance. Let alone the laws and regulations to consider for field testing.

Given these limitations and problems, lots of autonomous driving simulators and open-source prototypes are proposed for research and development purposes. dSPACE provides an end-to-end simulation environment for sensor data processing and scenario-based testing with RTMaps and VEOS~\cite{dspace}. The automated driving toolbox is Mathwork's software, which provides algorithms and tools for designing, simulating, and testing ADAS and autonomous driving systems~\cite{simulink-tool}. AVL DriveCube is a hardware-in-the-loop driving simulator designed for real vehicles with simulated environments~\cite{AVL-DriveCube}. In addition to these commercialized products, there are also open-source projects like CARLA and Gezabo for urban driving or robotics simulations~\cite{koenig2004design,dosovitskiy2017carla}. 

Another promising direction is to develop affordable research and development of autonomous driving platforms. Several experiment platforms are quite successful for indoor or low-speed scenarios. HydraOne is an open-source experimental platform for indoor autonomous driving, and it provides full-stack programmability for autonomous driving algorithms developers and system developers~\cite{wang2019hydraone}. DragonFly is another example that supports self-driving with a speed of fewer than 40 miles per hour and a price of less than \$40,000~\cite{perceptin}.

% \liangkai{Add AVL Drive-Cube, matlab Mathworks}

\subsection{Physical Worlds Coupling}

Autonomous driving is a typical cyber-physical system~\cite{goswami2012challenges}, where the computing systems and the physical world have to work closely and smoothly. With a human driver, the feeling of a driver is easily coupled with the vehicle control actions. For example, if the driver does not like the abrupt stop, he or she can step on the brake gradually. In autonomous driving, the control algorithm will determine the speed of braking and accelerating. We envision that different human feelings, coupled with complex traffic environment, bring an unprecedented challenge to the vehicle control in autonomous driving. Take the turning left as an example: how fast should the drive-by-wire system turn 90 degrees? An ideal vehicle control algorithm of turning left should consider many factors, such as the friction of road surface, vehicle's current speed, weather conditions, and the movement range, as well as human comfortableness, if possible. Cross-layer design and optimization among perception, control, vehicle dynamics, and  drive-by-wire systems might be a promising direction~\cite{lv2018driving}. 

\section{Conclusion}
\label{conclusion}
The recent proliferation of computing and communication technologies like machine learning, hardware acceleration, DSRC, C-V2X, and 5G has dramatically promoted autonomous driving vehicles. Complex computing systems are designed to leverage the sensors and computation devices to understand the traffic environments correctly in real-time. However, the early developed autonomous vehicles' fatalities arise from time to time, which reveals the big gap between the current computing system and the expected robust system for level-4/level-5 full autonomous driving. In this paper, we present the state-of-the-art computing systems for autonomous driving, including seven performance metrics, nine key technologies, and twelve challenges and opportunities to realize the vision of autonomous driving. We hope this paper will bring these challenges to the attention of both the computing and automotive communities.

\ifCLASSOPTIONcaptionsoff
  \newpage
\fi



% \newpage
\bibliographystyle{IEEEtran}
\bibliography{main}
\end{document}